\documentclass[12pt, nofootinbib]{revtex4-1}
\usepackage{amssymb,amsmath}

\usepackage{graphicx}

\newcommand{\beq}{\begin{equation}}
\newcommand{\eeq}{\end{equation}}
\newcommand{\bes}{\begin{equation*}}
\newcommand{\ees}{\end{equation*}}
\newcommand{\bea}{\begin{align}}
\newcommand{\ena}{\end{align}}

\newcommand{\bra}{\langle}
\newcommand{\ket}{\rangle}
\newcommand{\abs}[1]{\lvert#1\rvert}

\newcommand{\Real}{\mathrm{Re}}

\begin{document}

\title{Non-adiabaticity and improved back-reaction}
\author{Curtis Asplund $^\dagger$ and David Berenstein $^{\dagger \ddagger}$}
\affiliation{$^\dagger$ Department of Physics, University of California at Santa Barbara, CA 93106\\
$^\ddagger$ Institute for Advanced Study, School of Natural Science, Princeton, NJ 08540}

\begin{abstract}
We treat quantum back-reaction in time dependent processes for quantum field theory in various simplified models. The first example is a harmonic oscillator whose frequency depends on a second quantum variable $x$. Beginning with a classical analysis,
we show how using a particular canonical transformation
the system can be described by an improved adiabatic
expansion with a velocity dependent force for $x$. We find an instability at a critical velocity that prevents integrating out the oscillator degree of freedom in the new variables. We extend this calculation to the quantum system and to field theory and describe how to study fermions with similar techniques. 
Finally, we set up a model with an abrupt change in the oscillator whose quantum mechanics can be solved exactly so that one can study the effects of back-reaction of a fully non-adiabatic change in a controlled setting. We comment on applications of these general
results to the physics of D-branes, inflation, and black holes in AdS/CFT.
\end{abstract}

\maketitle

\section{Introduction}

Many problems in quantum field theory that involve the time evolution of a quantum system are very hard to solve. For example, thermalization of a system given some initial condition is such a problem and is crucial for 
understanding experiments like RHIC. Current methods in field theory do not explain the fast thermalization seen in this experiment \cite{Ackermann:2000tr}. 

As an approach to the problem of thermalization, one can study the excitation of degrees of freedom due to the time dependence of some other variables. Such problems typically arise in the study of cosmology, in particular the problem of reheating \cite{Kofman:1994rk}. Similar problems arise in string theory when one studies the collisions of D-branes and string creation between D-branes.

These problems usually start with some degrees of freedom that are in motion and some other degrees of freedom that become excited due to this motion. The first set (those that are initially in motion) are usually called the moduli fields. Let us call them $\phi$. During this motion there are other degrees of freedom, e.g. the modes of a massive field, that are sensitive to the motion of the moduli. Let us call these the heavy degrees of freedom. The heavy degrees of freedom have large associated frequencies $\omega(\phi)$, so these degrees of freedom are fast. In most field theory setups these can be thought of as harmonic oscillators.

 Let us suppose they all start in the ground state. If the moduli motion is classical and the velocity small, then the adiabatic theorem guarantees that they stay in the ground state \cite{Messiah:1958}. However, if the velocity of the moduli fields is large enough or if $\omega(\phi)$ small enough, there can be transitions to excited states. This results in particle production in the field theory case. This describes the effect of particle production in cosmological evolution (see \cite{BD} for a pedagogical example). In most cases, the moduli field motions are treated as a background on which one does computations for the heavy degrees of freedom, which simplify to a set of decoupled harmonic oscillators with $\omega(\phi(t))$ as time dependent frequencies. 
 
In such situations, the full calculation in the background field consists of a Bogoliubov transformation between the Fock space of heavy states in the initial time and the Fock space of heavy states in the final time \cite{Parker:1969au}. Here the particle production is considered to be small, so that interactions between the particles that are produced and the effects of these particles on the background moduli can be ignored. With some extra effort, dissipation can be added to the background moduli to account for some of the energy lost to the particles that are produced.

We are interested in a more systematic treatment of the production of particles with the back-reaction of the moduli included. We want to improve on the adiabatic approximation and to understand how it breaks down. This is particularly important in setups where one expects strong back-reaction due to the excitation of the heavy degrees of freedom, say due to fast thermalization. This mechanism 
is believed to operate in the problem of black hole formation in the dual CFT of AdS/CFT setups \cite{Asplund:2008xd} and this is one of the main motivations for this work. The black hole formation process has been well studied on the gravity side \cite{Bhattacharyya:2009uu} (and references therein), but the problem of the time evolution and thermalization of the quantum field theory is largely unsolved. However, see \cite{Iizuka:2008hg} for a solvable toy model that addresses some of these issues.

For all of these problems we want to study what happens once we take into account this particle production mechanism in the dynamics and include it in the evolution of the moduli fields. It is important to approach the problem with caution, for if we think of particle production as a measurement, the different particle production outcomes could decohere and each such state would then evolve independently of the others. With this in mind we do not average over the heavy degrees of freedom, as this might lead to wrong results. 

In this paper we begin to systematically explore these effects in simple toy model systems. We will have one modulus (light) field and one heavy field whose mass depends on the modulus field, treating the cases of heavy bosons and fermions separately. We will find it more convenient to work in the Hamiltonian formalism. The process of integrating out degrees of freedom in a background field is more common in the Lagrangian formalism in field theory applications. This same procedure of integrating out fast degrees of freedom is performed in the Born-Oppenheimer approximation in the study of molecules or solids, where there is also extensive work on higher-order corrections \cite{Baer:2006}. A lot of our work is to rewrite the Hamiltonian in variables that make this procedure more transparent and that permit one to study systematic improvements to the adiabatic approximation in a manner and setup appropriate to the applications we have in mind.

In the next section we will write a Hamiltonian for the light and heavy fields in a basis that diagonalizes the heavy field degrees of freedom for each value of the modulus, considering first the classical then the quantum problem. Here the appropriate adiabatic approximation is the Born-Oppenheimer approximation where the heavy fields are integrated out. We keep a parameter that controls how much back-reaction is present, the classical mass of the modulus degree of freedom. At high mass and fixed velocity, the kinetic term of this degree of freedom dominates the energy, so the force due to the heavy fields not being in their ground state has a negligible effect. That is, the degree of freedom has a lot of inertia. For low mass, any small modification of the potential has a large effect, because the kinetic energy is small. In the third section we discuss what happens in the case of heavy fermions. In the fourth and final section we set up a simplified model with an abrupt change in the oscillator degree of freedom whose quantum mechanics can be solved exactly and we investigate and interpret the solutions.

\section{The minimal model}

	We begin with a simple quantum system with two coupled degrees of freedom, one of which can interpreted as a rolling modulus field. This is a minimal model for studying back-reaction in the regime beyond which the adiabatic approximation breaks down: we need one adiabatic degree of freedom that turns non-adiabatic and we also need another degree of freedom on which to back-react. 
	Our model is specified by the following quantum Hamiltonian
\beq
\label{H}
	H(x,y) =  -\frac{1}{2m_x}\frac{\partial^2}{\partial x^2} - \frac{1}{2} \frac{\partial^2}{\partial y^2} + \frac{1}{2}\Omega^2(x) y^2 .
\eeq
This is the Hamiltonian for a particle in two dimensions with potential $U(x,y) =  \frac{1}{2}\Omega^2(x) y^2$ (in units where $\hbar = 1$). We normalize the mass of $y$ to unity but keep the mass parameter $m_x$ of $x$. The function $\Omega$ controls the curvature of the quadratic potential well in the $y$ direction. A similar toy model was considered in the problem of D-brane scattering \cite{Douglas:1996yp}, where $\Omega(x) \propto \abs{x}$. A classical analysis of this simple model can be found in \cite{Helling:2000kz}.

The case of a harmonic oscillator in the $y$-direction with angular frequency $\omega$ is just $\Omega(x) \equiv \omega$.
In that case the Schr\"{o}dinger equation may be easily solved (via separation of variables), yielding the spectrum of eigenfunctions with wavefunctions
\beq
\label{SHOsoln}
	 \psi_{k,n}(x,y) = e^{ikx} N_n H_n(\omega^{1/2}y) e^{-y^2\omega/2}
\eeq
where $N_n = (\omega/\pi)^{1/4}\left(1/\sqrt{n! 2^n} \right)$ is a normalization constant and $H_n$ is the $n^\text{th}$ Hermite polynomial. Here $k\in \mathbb{R}$ and $n \in \mathbb{N}$. The corresponding energies are 
\begin{equation}
E_{k,n} = k^2/2m_x + \omega(n + 1/2).
\end{equation}

In a classical setup we can always set $y=p_y=0$ for any $\Omega(x)$ and get a solution of the equations of motion. The $x$ variable then behaves as a free particle and the solution to the equations of motion is $x=vt$, where $v$ is some velocity. Thus $x$ essentially measures the time. We can even consider a classical potential for $x$ if we want to, but the point of our paper is to understand $y$ and how this degree of freedom back-reacts onto the motion in $x$, especially in the quantum system, so we will not study the dependence on a potential in $x$.

Assuming $x=vt$, we can then study infinitesimal variations around this solution with $y$ and $p_y$ small. As such, the $y$ and $p_y$ infinitesimal motions represent a harmonic oscillator with a time dependent frequency, given by $\omega(t)=\Omega(vt)$. The motion in these variables is considered adiabatic if 
\begin{equation}
\frac{\dot \omega}{\omega^2}= v\frac{ \Omega'}{\Omega^2} \ll 1.
\end{equation}
This can be achieved with small velocity, large $\Omega$ or a small gradient of $\Omega$. But if we increase $v$ enough, we can always go to a velocity regime where the adiabatic approximation breaks down. We will see that it is exactly in this regime that we can not ignore quantum corrections. 

We would like to understand the breakdown of adiabaticity of this system, in the form of a variation of $\Omega(x)$. That is, we would ultimately like to have an expansion (of, say, the Hamiltonian) in terms of derivatives of $\Omega$ that lets us systematically correct adiabatic or Born-Oppenheimer approximations and lets us know when such approximations break down entirely. Remember also that in a quantum system $\dot \Omega(x)$ gets replaced by an operator, so the use of $x$ as time becomes more of an issue.

In standard semiclassical treatments, one would treat $x$ classically but solve for $y$ quantum mechanically. Motion in $x$ can lead to particle production (excitation) in the $y$ direction. The equations of motion of $x$ can then be corrected by averaging over the particle production of $y$. However, as we mentioned in the introduction, this averaging procedure can be problematic for solving the real quantum problem accurately, so we do not do this here.

\subsection{Breakdown of adiabaticity}

	We can vividly see the breakdown of adiabaticity by first considering the system classically. The classical limit of this quantum system is given by
\begin{equation}
H_{\mathrm{class}}=\frac {p_x^2}{2m_x}+\frac {p_y^2}{2} +\frac 12 \Omega^2(x) y^2.
\end{equation}
The appearance of $y$ in the solutions \eqref{SHOsoln} always as $\omega^{1/2}y$ suggests that a change of variables $\tilde{y} = \Omega^{1/2} y$ may be useful in understanding the essential degrees of freedom in the system. So let us consider the system in terms of the variables $\tilde y$ and $\tilde x= x$. We have that
\begin{align}
\partial_y &= \frac{\partial \tilde y}{\partial y}\partial_{\tilde y}+\frac{\partial\tilde x}{\partial y}\partial_{\tilde x}= \Omega^{1/2}(\tilde x) \partial_{\tilde y}\\
\partial_x &= \frac{\partial \tilde y}{\partial x}\partial_{\tilde y}+\frac{\partial\tilde x}{\partial x}\partial_{\tilde x}= \partial_{\tilde x} +\frac 12\frac {
\Omega'(\tilde x)}{ \Omega(\tilde x)} \tilde y \partial_{\tilde y}. 
\end{align}
These can then be substituted into the Hamiltonian. There is also a change of measure $d\tilde x d\tilde y= \Omega^{1/2} dx dy$, 
and one has to be careful about this. We will address this in the quantum case in the next subsection.

To implement this change of variables in phase space, we make the canonical transformation
\begin{align}
	\tilde{x} &:= x \\
	\tilde{y} &:= \Omega^{1/2}(x) y \\
	p_{\tilde{x}} &:= p_x - \frac{1}{2} \frac{\Omega'(x)}{\Omega(x)} y p_y \\
	p_{\tilde{y}} &:= \frac{p_y}{\Omega^{1/2}(x)}.
\end{align}
Canonicity of this transformation can be checked from the invariance of the Poisson brackets. Notice that the change of variables in $y,p_y \to \tilde{y}, p_{\tilde{y}}$
is obviously a change of scale that is $x$ dependent and that their Poisson brackets are retained. Also, the Poisson bracket with $x=\tilde x$ vanishes as before. However, $p_x$ does not Poisson-commute with $\tilde y,  p_ {\tilde{y}} $ and it gets corrected. This can be guessed by noticing that the bracket of $p_x$ with $\tilde y$ is proportional to $\tilde y$, and the bracket of $y p_y$ with $\tilde y$ is also proportional to $\tilde y$. Therefore we can try to cancel these against each other, getting the result above. There will be a similar situation when we deal with fermions.

The Hamiltonian in these new variables is
\beq
	\label{Hamil2}
	H =   \frac{1}{2m_x}p_{\tilde{x}}^2 + \frac{\Omega}{2} (p_{\tilde{y}}^2+\tilde{y}^2) + \frac{1}{2m_x}\frac{\Omega'}{\Omega} \tilde{y} p_{\tilde{y}} p_{\tilde{x}} + \frac{1}{8m_x} \left(\frac{\Omega'}{\Omega} \right)^2 \tilde{y}^2 p_{\tilde{y}}^2  ,
\eeq
where $\Omega = \Omega(\tilde{x})$ and $\Omega' = \partial \Omega(\tilde{x})/ \partial \tilde{x}$. Notice that the Hamiltonian receives both quadratic and non-quadratic corrections in the $\tilde y, p_ {\tilde{y}} $ variables. 
 The coefficient of the former is given by 
\begin{equation}
\frac{p_{\tilde{x}}}{2m_x}\frac{\Omega'}{\Omega} \simeq \frac 12 \frac{\dot \Omega}{\Omega},
\end{equation}
the latter expression valid in the $\tilde{y} = p_{\tilde y}=0$,  $p_{\tilde{x}}/m_x= v$ regime. 

\begin{figure}[ht]
\centering
	\includegraphics[width=2in]{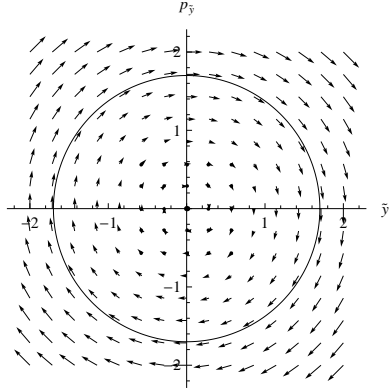}
	\caption{The phase portrait of $\tilde{y}$ and $p_{\tilde{y}}$ for constant $\Omega$}
	\label{circle1}
\end{figure}

	From this Hamiltonian we can already see that there will be a qualitative change in the behavior of the $\tilde{y}$ degree of freedom as $\Omega'$ differs from zero.  For constant $\Omega$, the Hamiltonian for $\tilde{y}$ and $p_{\tilde{y}}$ is simply a classical harmonic oscillator, with phase portrait as shown in Figure \ref{circle1}.

\begin{figure}[ht]
\centering
	\includegraphics[width=2in]{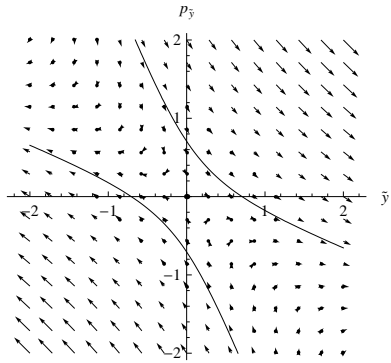}
	\caption{The phase portrait of $\tilde{y}$ and $p_{\tilde{y}}$ for non-negligible $\Omega'/\Omega$}
	\label{hyperbola1}
\end{figure}

	As $v \Omega'/\Omega$ becomes non-trivial, the third and fourth terms in \eqref{Hamil2} become important, changing the orbits from circular first to elliptical and then to hyperbolic. At that point the origin becomes an unstable fixed point as shown in Figure \ref{hyperbola1}. It is important to remember that here we are taking slices of phase space in the  $\tilde x, \tilde y$ variables and not the original $x,y$ variables of the system. 

We can be more precise. The behavior of the Hamiltonian vector field near the origin of phase space of $\tilde{y}$ is determined by the second degree terms in $\tilde{y}$ and $p_{\tilde{y}}$ in \eqref{Hamil2}. The stability of the solutions is set by the signs of the eigenvalues of the quadratic form
\beq
{\bf \Omega}= 2 \begin{pmatrix} \Omega/2 & (\Omega'/\Omega) p_{\tilde{x}}/4m_x \\ (\Omega'/\Omega) p_{\tilde{x}}/4m_x & \Omega/2 \end{pmatrix}
\eeq
(the overall factor of 2 is added for later convenience). The condition that the origin is stable is that the determinant of this matrix is positive ($\Omega$ is positive in our conventions). This can be seen by diagonalizing, which puts the Hamiltonian in the form of a simple harmonic oscillator and the sign of the determinant then corresponds to the sign of the quadratic potential. The condition for stability then gives us
\beq
\label{adiabatic1}
	\left\vert \frac{\Omega'}{\Omega^2}\frac{p_{\tilde{x}}}{2m_x} \right\vert  < 1 .
\eeq
This is analogous to the non-adiabaticity condition $\dot{\omega}/\omega^2 \lesssim 1$, often used as a threshold for non-adiabatic behavior or particle production. Here it sets the threshold for treating $\tilde{y}, p_{\tilde{y}}$ as a harmonic oscillator versus an oscillator with an inverted potential (this requires rotations
of the variables to make it explicit). That is, when we have a classical motion in $x$ the $\tilde y$ degree of freedom becomes unstable exactly when
\begin{equation}
\frac{\dot \omega}{\omega^2}=2 
\end{equation}
and we expect to have corrections of order one in the $\tilde y$ degree of freedom, when ${\dot \omega}/{\omega^2} \simeq 1$. Indeed, we can improve the adiabatic approximation with the diagonalized quadratic action in $\tilde y, p_{\tilde{y}}$. Recalling that the canonical form of the harmonic oscillator is the diagonal form $p^2/2m + m\omega^2 x^2/2$, the diagonalization gives us an oscillator with a (position and momentum dependent) frequency
\begin{equation}
\tilde \omega =  \sqrt{\det {  \bf\Omega}} =  \sqrt{\Omega^2- \frac {p^2_{\tilde x}}{4 m_x^2}\left(\frac{\Omega'}{\Omega}\right)^2}. \label{eq:improved}
\end{equation}

If we consider a semiclassical treatment with $x = vt$ to quadratic order in $y$, the zero point energy of such an oscillator is then
\begin{equation}
\frac 12 \tilde \omega=  \frac 12  \sqrt{\Omega^2- \frac 1{4}\left(\frac{\dot\Omega}{\Omega}\right)^2} \simeq 
\frac 12 \Omega -\frac 1{16} \Omega \left( \frac{\dot\Omega}{\Omega^2}\right)^2+ \dots \label{eq:taylor},
\end{equation}
where the last expression holds for small $\dot{\Omega}/\Omega^2$ and we have used $p_{\tilde{x}}/m_x \simeq v$ and $\dot\Omega = \Omega' v$. We see that in this way it is possible to generate velocity dependence in the low-energy effective Hamlitonian for the moduli degrees of freedom, in this case $\frac{1}{2}(p_{\tilde{x}}^2/m_x + \tilde{\omega})$. Notice that there is apparently already an $x$-velocity dependence in the term in \eqref{Hamil2} linear in $p_{\tilde{x}}$. However, when integrating out $\tilde y$ and $p_{\tilde y}$ one finds that to first order in $\dot\Omega$ this term has no contribution since $\tilde{y} p_{\tilde y}$ averages to zero over the motion in this regime. Thus this term should contribute only at second or higher order, consistent with our result above.

Velocity dependent terms like this have been argued to give rise to gravitational interactions in matrix theory as a dual description of M-theory in 11-D \cite{BFSS}. Also notice that the square root formula is reminiscent of a Dirac-Born-Infeld Hamiltonian for D-branes, but the sign of the velocity term in the square root is wrong for that comparison. As we mention in the conclusion, a full investigation of these issues and connections is reserved for future work.

When the origin in the $\tilde y, p_{\tilde y}$ plane becomes unstable, the fact that in a quantum system a state occupies some finite area of phase space implies that the
quantum state will spread along the trajectory of instability. At that stage we cannot integrate out the $\tilde y$ variable any longer and it has to participate in the full dynamics. This is a nice semiclassical way to explain that there  must be particle production at large $v$ compared to the case of $v$ near zero. In the following sections we will turn to the quantum system and explain how to treat the adiabatic approximation and its breakdown more carefully.

We can interpret the effect on the $x$ degree of freedom as a momentum dependent force \footnote{Such terms are usually computed as a correction to the action in the path integral formulation around a classical solution of the $x$ motion. This is exemplified in D-brane setups in \cite{Bachas:1995kx}, where one also sees the appearance of particle production from the imaginary part of the phase shift.  However, no back-reaction is taken into account.}.  Indeed, let us generalize the above result and consider a case where the degree of freedom $x$ is some rolling modulus field and we have field theory degrees of freedom $y_k$ labeled by their momenta in a $d$-dimensional box of volume $V$.
The corresponding field theory would live in $d+1$ dimensions.

 The effect we consider will just sum over the modes. 
 For each of them we will find the usual relativistic dispersion relation, where $\Omega(x)$ plays the role of the mass. The frequencies for the mode labeled by the momentum $k$ is given by
 \begin{equation}
\Omega_k= \sqrt{\Omega^2+k^2}
\end{equation}
and thus
\begin{equation}
\Omega'_k= \frac{\Omega\Omega'}{\sqrt{\Omega^2+k^2}}.
\end{equation}
We find this way that the modified frequencies \eqref{eq:improved} are given by
\begin{equation}
\tilde \omega_k= \sqrt{\Omega^2+k^2 -\frac 1{4} \left(\frac{\Omega \dot \Omega}{\Omega^2+k^2}\right)^2 }.
\end{equation}

The zero point energy contribution (after subtraction of the term at zero velocity in order to get a finite answer) gives us an additional contribution
\begin{equation}
\frac 12 \sum_k :\tilde{\omega}_k:\ =   \sum_k \frac 12 \sqrt{\Omega_k^2- \frac 1{4}\left(\frac{\dot\Omega_k}{\Omega_k}\right)^2} -\frac12\Omega_k \simeq 
-\sum_k \frac 1{16}\Omega^{-5}_k{(\dot\Omega \Omega)^2}+ \dots\label{eq:taylor2}.
\end{equation}
The sum is convergent if the number of spatial dimensions is less than five, $d<5$. The effective term in the hamiltonian, after integrating over all modes, is proportional to  
\begin{equation}
\label{eq:allmodes}
- V\frac{(\dot \Omega)^2}{\Omega^{3-d}} 
\end{equation}
where $V$ is the volume of the $d$ dimensional box: it fixes the dimensional analysis in the equation. We find this way a finite effect per unit volume, with a fixed sign. In three spatial dimensions $d=3$ we notice that the correction only depends on the gradient of the mass and the velocity of the modulus field. This can be generalized to multi-field moduli, giving essentially the same result, where now $\dot \Omega= \nabla_i \Omega \pi_i/m$ and $\pi_i$ is the canonical conjugate of $x^i$. This first term in the correction can be interpreted then as a modified metric on the moduli space (a sigma model would be written with a Hamiltonian given by $H = g^{ij} \pi_i \pi_j$). It has a definite sign, making distances longer along the gradient of $\Omega$ (the sign is negative above because it is associated with the inverse metric) and giving rise to a curved metric on the moduli space.

\subsection{Quantum adiabatic approximation and corrections}

Now that we have described the classical theory with a brief excursion into a simplified description of the quantization, we turn our attention to dealing with the fully quantum problem. Within the adiabatic regime where \eqref{adiabatic1} is satisfied we are motivated by the above canonical transformation to consider wavefunctions of the form
\beq
\label{wavefunc2}
	\psi(x,y) = \sum_{n=0}^\infty f_n(x) \left(\frac{\Omega(x)}{\pi}\right)^{1/4} \frac{1}{\sqrt{n! 2^n}} H_n(\Omega^{1/2}(x) y) e^{-y^2\Omega/2},
\eeq
and so specify a state by the set of functions $\{f_n(x)\}$. This is a change of basis or representation in the Hilbert space of states. There is no loss of generality from writing the wave function in this fashion and the following results are still exact.

Wavefunctions of this form have the nice property of being energy eigenstates with respect to the $y$ parts of the Hamiltonian (at each $x$). This setup and the following analysis is exactly the Born-Oppenheimer method of solving the dynamics of quantum systems. One first divides the variables into slow and fast degrees of freedom. One then solves the exact dynamics of the fast degrees of freedom when the slow degrees of freedom are frozen and compute the Hamiltonian in the new basis. The further assumption to simplify the problem (the Born-Oppenheimer approximation) is that no transitions occur between different levels as the slow degrees of freedom move, mainly that the motion is adiabatic. This approximation has been studied extensively. We would like to draw particular attention to the works of M. Berry
\cite{Berry:1984jv} where this setup was explored systematically. Apart from the Berry phase connection, this generally leads to effective terms in the theory that can   be attributed to effective electric fields and magnetic fields for the slow variables (see \cite{BerryRob} and references therein) \footnote{Many of the standard results of D-brane interactions  \cite{Pol}  due to integrating open strings can be rephrased in this language.}. Our results do not assume the adiabatic approximation, but we do perform the appropriate adiabatic change of variables.

  We have chosen the wave functions for $y$ to be normalized to one at fixed $x$. This means that the change of measure from the $x,y$ variables to the $\tilde x= x, \tilde y = y\Omega^{1/2}$ has been absorbed in the normalization of these states, which is why we encounter $\Omega(x)^{1/4}$ sitting in front of the $y$ wave functions. This implements the quantum change of variables so that the 
$y$ part of the wave function has the correct behavior with respect to $\tilde y$ variables in the number occupation basis for the oscillator of frequency 
$\Omega(x)$. 

We now compute the matrix elements of the Hamiltonian in this basis of states. Non-adiabatic changes in $\Omega$ will manifest as off-diagonal contributions to the Hamiltonian in this basis, which is what we're interested in. In a general energy eigenstate basis $\{ |\psi_i \ket \}$ we can act on a state $|\psi\ket = \sum_n c_n |\psi_n \ket$  with $H$ to obtain $H|\psi\ket = \sum_n E_n c_n |\psi_n \ket$. That is, we obtain a new state with coefficients weighted by the corresponding energies. Here we do an $x$ dependent version of this, which will tell us how the Hamiltonian mixes states of the form \eqref{wavefunc2}.  

We write the wavefunction of $H|\psi\ket$ in the same form as \eqref{wavefunc2}, with a new set of functions $\{g_m(x)\}$. Then we can extract a matrix representation of the Hamiltonian by using orthonormality and integrating out the $y$ variable:
\beq
	g_m(x) = \sum_n H_{mn}(x) f_n(x)
\eeq
where (no summation on $n$)
\bea
	H_{mn}f_n &= \frac{1}{\sqrt{m!2^m}}\frac{1}{\sqrt{n!2^n}} \left( \frac{\Omega}{\pi} \right)^{1/4} \int\ H_m(\Omega^{1/2}(x) y)\ e^{-y^2\Omega/2} \\ \nonumber
	&\times \left[ -\frac{1}{2m_x}\frac{\partial^2}{\partial x^2} - \frac{1}{2}\frac{\partial^2}{\partial y^2} + \frac{y^2}{2}\Omega^2(x)\right] f_{n}(x) \left(\frac{\Omega(x)}{\pi}\right)^{1/4}  H_{n}(\Omega^{1/2}(x) y)\ e^{-y^2\Omega/2}\ dy.
\end{align}
One term in the diagonal part of this matrix can be computed immediately, since for fixed $x$ the factors $H_n$ and the exponential are an eigenfunction of $-\partial_y^2/2 + \Omega^2y^2/2$, with eigenvalue $\Omega(n+1/2)$. So along with the orthogonality of the Hermite polynomials we have
\beq
	H_{mn} f_n = f_m \Omega (m+1/2) \delta_{mn} + (\text{integral with -$\partial_x^2$}/2m_x).
\eeq

The computation of the remaining integral is straightforward but long, as the second derivative gives rise to 16 terms. But each can be evaluated using the orthogonality and recursive properties of the Hermite polynomials. The result is 
\begin{align}
	H_{mn} f_n = &- \frac{1}{32 m_x} \sqrt{(m+1)(m+2)(m+3)(m+4)}\ f_{m+4}\ \frac{\Omega'^2}{\Omega^2} \delta_{m(n-4)} \nonumber  \\
	&- \frac{\sqrt{(m+1)(m+2)}}{8m_x} \left[f_{m+2} \left(\frac{\Omega''}{\Omega} - \frac{\Omega'^2}{\Omega^2}\right)  + 2f'_{m+2}\frac{\Omega'}{\Omega} \right] \delta_{m(n-2)} \nonumber \\
	&+ \left[f_m \Omega (m+1/2) - \frac{f_m''}{2m_x} + f_m \frac{\Omega'^2}{\Omega^2} \frac{m^2 + m + 1}{16 m_x}\right] \delta_{mn} \nonumber\\
	&+ \frac{\sqrt{m(m-1)}}{8m_x} \left[f_{m-2} \left(\frac{\Omega''}{\Omega} - \frac{\Omega'^2}{\Omega^2}\right)  + 2f'_{m-2}\frac{\Omega'}{\Omega} \right] \delta_{m(n+2)} \nonumber\\
	&- \frac{1}{32m_x} \sqrt{m(m-1)(m-2)(m-3)}\ f_{m-4}\ \frac{\Omega'^2}{\Omega^2} \delta_{m(n+4)}.
\end{align}
Here we have an explicit expansion of $H$ in terms of the derivatives of $\Omega$ \footnote{This is the same expansion as in the Born-Oppenheimer method, where the fast degrees of freedom are oscillators rather than electrons in a molecule or material. The off-diagonal terms correspond to the ``vibronic couplings" or ``non-adiabatic coupling terms" in that context \cite{Baer:2006}. Our results in this section are a specialization of the Born-Oppenheimer method to a case where the fast degrees of freedom are oscillators, which can be interpreted as modes of a massive quantum field.}. The adiabatic approximation in this context is to ignore all terms in the Hamiltonian that depend on derivatives of $\Omega$. Notice that these are the only terms that can change the level of the oscillator, so in the adiabatic approximation  we end up projecting onto the components of $H$ that lie on the diagonal $H_{mn}\propto \delta_{mn}$, giving the result

\begin{equation}
H^\mathrm{ad}_{mn} f_n =  \left(f_m \Omega (m+1/2) - \frac{f_m''}{2m_x}\right).
\end{equation}	
The system splits into an infinite number of one dimensional problems for a particle on a line with potentials 
\begin{equation}
V_m(x)= \Omega(x) \left(m+\frac 12\right).
\end{equation}
These define what are called the potential energy surfaces, with a conical intersection at $x=0$, in the Born-Oppenheimer literature. We see here the usual energy of a harmonic oscillator in the $y$ variable including the zero point energy.

We can now expand on the terms that are ignored in an adiabatic approximation as a perturbation of the Hamiltonian. 
The terms that involve $x$-derivatives of the wave function coefficients $f_n$ only change the level by $\pm 2$ units (in particular, they keep the even/odd splitting of the $y$ wave functions). The terms that do not involve $x$-derivatives of the $f_n$ involve two $x$-derivatives acting on polynomials of $\Omega$. Only the first set of terms,  involving an $f'_n$,  are momentum dependent. If we compare these with our classical result given in equation \eqref{Hamil2}, we find similarities in the description.
In terms of lowering and raising operators, we usually have that $\tilde y \propto a+a^\dagger$, while $ p_{\tilde y}\propto a-a^\dagger$. The term $\tilde y p_ {\tilde y}$ is then of schematic form $a^2 - (a^\dagger)^2$, up to normal ordering ambiguities. A careful analysis shows that the off-diagonal terms that involve derivatives of the form $f'_m$ exactly match the harmonic oscillator algebra coefficients. 
Thus, we can state that in the quantum theory, the momentum dependent corrections depend on the operator 
\begin{equation}
\frac{\Omega'}{\Omega}\frac{p_x}{4 m_x} (i (a^\dagger)^2-i a^2).
\end{equation}

For semiclassical $x$ motion, $p_x/m_x\simeq v$ and if we take into account only these corrections, we can obtain a power series in $v$ for the ground state. This arises from the self-consistent assumption $f_n \simeq O(v^n)$, beginning with $f_0 \simeq O(1)$ (but this can be generalized to other states). The corrections to the energy etc, will then always be even powers of the velocity. These terms survive in the $m_x\to \infty$, $v$ fixed limit.
This limit is the classical limit for $x$. At large mass for the motion in $x$, fixed $\Omega(x)$, there is essentially no back-reaction. This is because the system is dominated by the kinetic energy $E\simeq \frac 12 m_x v^2$ and $\Omega$ is a parametrically small perturbation. Also notice that the terms that do not involve momentum are those that correspond to the square $(\tilde y p_{\tilde y})^2$ and these are suppressed in the large 
$m_x$ limit. 

We can also work in a level truncation approximation (again removing the terms with no $f'$), where we keep only the modes $n=0, n=2$. The Hamiltonian in two components will then be of the form
\begin{equation}
H\begin{pmatrix} f_2\\
f_0\end{pmatrix} =\left( -\frac 1{2m_x}\partial_x^2 +\Omega(x)\begin{pmatrix}
 5/2 &0\\
0&1/2\end{pmatrix}+i  \frac{\sqrt 2 \Omega'}{4 m_x \Omega}\begin{pmatrix} 0& i\partial_x\\
-i \partial_x &0 \end{pmatrix}\right)\begin{pmatrix} f_2\\
f_0\end{pmatrix}. \label{eq:2level}
\end{equation}
If we set $f_0 \simeq c_0 \exp(ikx)$ and $f_2\simeq c_2 \exp (ikx)$ to check how important the off-diagonal terms are (again in a limit where $x$ is essentially classical), we have a level splitting of $2 \Omega$ compared with the off-diagonal component that is $\sqrt 2 \Omega'/\Omega \frac{ k}{4m_x}$.

Diagonalizing the above Hamiltonian we find that the lowest energy level for the two-level system is
\begin{equation}
\frac 32 \Omega - \sqrt{ \Omega^2+ \frac 18\left(\dot\Omega/\Omega\right)^2  } \simeq \frac 12 \Omega-\frac {1}{16}\Omega \left(\frac{\dot\Omega}{\Omega^2}\right)^2 + \ldots.
\end{equation}
We see that to second order in the Taylor series in $\dot\Omega$ we find a match with our semiclassical analysis in \eqref{eq:taylor}, making the naive semiclassical result more plausible.

One should notice that the sign of the term that includes $\dot \Omega^2$ is negative. This is expected because the perturbation is off-diagonal and under this condition in second order perturbation theory in quantum mechanics the correction to the energy of the ground state is non-positive.

If we go beyond the adiabatic regime and pass through a region where the $\tilde y$ variable is unstable, the semiclassical result \eqref{eq:taylor} becomes purely imaginary in this region. The integrated effect of this term can be interpreted as an imaginary contribution to the action for a persistence amplitude in the ground state, indicating that the system ends up in a state different than the ground state with some finite probability.

It's useful to estimate the probability that we go from the level with zero occupation number to the level with occupation number two in the regime where there is no back-reaction on $x$. Remember that having no back-reaction on $x$ is equivalent to the limit where we send the mass of the $x$ degrees of freedom to infinity, keeping the velocity fixed. Also, $x(t)$ becomes a classical trajectory: at finite velocity the momentum conjugate to $x$ is very large and we can localize the wavepacket in the $x$ direction to arbitrary precision without having to take into account the finite time broadening of the wave packet.

In such a system, we want to understand the dynamics of the $y$ degree of freedom. For simplicity,
 we can truncate the system to two levels as above, in the regime where the off-diagonal terms are parametrically small (the truncation to these levels is only valid in this regime). Then $\Omega(x) \simeq \Omega(x(t))$ and the system is a two level system with effective Hamiltonian given by
\begin{equation}
H(t) = \begin{pmatrix} 2\Omega +\frac 12 \Omega& \sqrt 2 i  \frac{\dot \Omega}{4\Omega}\\
- \sqrt 2i   \frac{\dot\Omega}{4\Omega}& \frac 12 \Omega
\end{pmatrix}.
\end{equation}
In such a system the off-diagonal terms constitute the perturbation Hamiltonian. The difference in energy between the excited state and the unexcited state means the system will be subject to Rabi Oscillations. We will see that we get a small average occupation number if the off-diagonal term is parametrically small, i.e. the system is near adiabatic. 

The amplitude is given to first order in time dependent perturbation theory by
\begin{equation}
A_{0\rightarrow 2} =  \frac{i}{2\sqrt 2} \int_0^T dt\ \frac{\dot \Omega}{\Omega} \exp \left( i \int_0^t ds\ 2 \Omega(x(s))  \right). 
\end{equation}
To estimate, we take $\Omega$ large, $\dot \Omega$ small and both approximately constant over a time $T$. This is then roughly given by
\begin{equation}
A_{0\rightarrow 2} \simeq \frac i {2\sqrt2} \frac{\dot \Omega}{\Omega} \int_0^T dt \exp( 2 i \Omega t)= \frac i{2\sqrt2} \frac{\dot \Omega}{\Omega^2} \exp(i\Omega T ) \sin(\Omega T).
\end{equation}
The probability of finding the state in the excited configuration will then be given by 
\begin{equation}
\label{eq:Rabiprob}
P(T) \simeq \frac{\eta^2}{8}\sin^2(\Omega T),
\end{equation}
where $\eta = \dot \Omega/ \Omega^2$ is the adiabaticity parameter. 

This averages over time to $\eta^2/16$. So we find that the probability of being excited grows like the velocity squared, but stays small in the adiabatic regime $\eta \ll 1$.
Notice that this is a short time estimate (we assumed $\Omega$ approximately constant).  We can do a similar calculation if we already begin in an excited state at level $m$ and consider transitions $m\to m\pm 2$. We find that the amplitude to excitation is slightly larger than to de-excitation because of the harmonic oscillator algebra. This is similar to stimulated emission of radiation: we get signal amplification. This is typical for this class of problems \cite{Parker:1969au}.

Notice that this is very different from particle production for an infinitely long time process. For many of these the amplitude over an infinite regime becomes  
non-perturbatively suppressed in the velocity for small velocity, so long as there is a minimum frequency in $\Omega(t)$. This is typical of D-brane scattering with finite impact parameter. This is nicely explained in \cite{Kofman:2004yc}.

The main reason for extra suppressions at large times is the following. At constant velocity for $x$ we have
\begin{equation}
\Omega(x(t))= \Omega(vt).
\end{equation}
Then $\dot \Omega/\Omega^2 =  v \Omega'/\Omega^2$ and the phase in the exponential is given by
\begin{equation}
\int_0^t ds\ 2 \Omega(v s) = \frac{1}{v} \int_0^{x = vt} du\ 2 \Omega(u).
\end{equation}
The transition amplitude takes the form 
\begin{equation}
A_{0\to 2}= \frac{i}{2\sqrt{2}} \int_{-\infty}^\infty dx \frac{\Omega'(x)}{\Omega^2} \exp\left(\frac i v \int_0^x du\ 2\Omega(u) \right), 
\end{equation}
which oscillates very quickly when we send $v\to 0$ and the behavior of the integral is similar to that of the Fourier transform of $\Omega'/\Omega^2$ at high frequency.
We have in mind changing variables to $\gamma=  \int_0^x \Omega(u) du$. 
The resulting integral receives contributions from the poles in the complex plane where $\Omega(x(\gamma))=0$ and this can be very different from integrals at finite time where we cannot deform the contour.

Notice that this number is generally much smaller than $A_{0\to 2}$ for finite time, where the result is polynomial in the velocity. The result that is polynomial in the velocity contributes to back-reaction as a power series in $v$, matching the type of expansion we saw before and what is expected from direct calculations of velocity dependent forces.

\section{Fermions}
\label{fermions}

We now discuss fermions for completeness, both at the semiclassical level and at the quantum level. We consider systems where there is a fermion parity operator $(-1)^F$, with $F$ counting the number of fermions, that describes whether a state is fermionic or bosonic and is preserved by the action of the Hamiltonian. Since we wish to study a system that can mix with a bosonic system, we need at least two different fermions, giving us at least four states: two of even parity, two of odd parity. The even states can mix with each other, but not with those of odd parity. To follow the discussion of bosons above we will need to use the Hamiltonian formalism for fermionic (anticommuting) variables, as described below. We begin with a single fermion oscillator then consider the possibilities when additional fermions are present.

A fermion oscillator is defined with a pair of Grassmann variables $\theta^{1,2}$ with Lagrangian given by 
\begin{equation}
L = \frac{1}2\left(  i \theta^{1} \dot\theta^1 + i \theta^{2} \dot\theta^{2} -  M_{ij} \theta^{i}\theta^{j}  \right).
\end{equation} 
The matrix $M_{ij}$ is antisymmetric, and the classical variables anticommute with each other:  $\{\theta^{i},\theta^{j}\}=0$. The $\theta$ variables are real:
$\theta^{i*}= \theta^{i}$. The factors of $i$ in the kinetic terms are there to make the Lagrangian real (using the usual property of complex conjugation that also reverses the order of the variables). The Hamiltonian is given by $\frac{1}{2} M_{ij} \theta^{i}\theta^{j}  $ and it is real if $M$ is antisymmetric and hermitean. 
The (left) canonical conjugates of $\theta^{1}, \theta^{2}$ are $\pi_1 = -i\theta^{1}/2$, $\pi_2= -i\theta^{2}/2$, and the Poisson brackets are
\begin{equation}
 \{ \pi_{i}, \theta^{j}\}_\textrm{PB} = -\delta_{i}^{j}.
\end{equation}
Here we are following the ``superclassical'' formalism and conventions of \cite{Henneaux:1992ig}. The equations of motion that follow from the Lagragian are
\begin{equation}
i \begin{pmatrix}\dot \theta^{1}\\
\dot \theta^{2}
\end{pmatrix}=  \begin{pmatrix} 0 & M_{12}\\
- M_{12}&0
\end{pmatrix}\begin{pmatrix} \theta^{1}\\
\theta^{2}
\end{pmatrix},
\end{equation}
corresponding to an oscillator with frequency $\omega = \abs{M_{12}} = \sqrt{\det M}$. For a detailed solution and more on the fermion oscillator see \cite{DeWitt:1992cy}.


	Now we consider coupling many such fermions to a classical variable $x$, so that $M_{ij}(x)$ is an $2N\times 2N$ antisymmetric, hermitean matrix that depends on $x$. The trick to solving the fermions is to rotate all the variables $\tilde \theta^{i} = R^{i}_{j} \theta^{j}$  into each other with an $\mathrm{SO}(2N)$ transformation so that $M$ takes a standard form
\begin{equation}
\tilde M(x) = R^T M R = \begin{pmatrix} 0&\tilde M_{12}(x)& 0 &0 &\dots\\
-\tilde M_{12}(x) &0&0&0&\dots\\
0&0&0&\tilde M_{34}(x) &\dots\\
0&0&-\tilde M_{34} (x) &0&\dots\\
\vdots &\vdots &\vdots&\vdots&\ddots
\end{pmatrix}.
\end{equation}

The Poisson bracket of $x$ with $\tilde \theta^{i}$ vanishes, but the Poisson bracket of $p_x$ with $\tilde \theta^{i}$ does not, analogous to the classical bosonic system we first considered. Indeed, we find that 
\begin{equation}
\{p_x, \tilde \theta^{i} \}_\textrm{PB}= -\partial_x R_j^i \theta^{j} =  -\partial_ x R_j^i (R^{-1})_k^j \tilde \theta^{k}.
\end{equation}
Now, we can correct $p_x$ by defining
\begin{equation}
 p_{\tilde x} = p_x + S^{i}_{j} \pi_i \theta^{j} 
\end{equation}
so that the Poisson bracket with $\tilde \theta^{i}$ vanishes. We find that $S = R^{-1} \partial_x R$.  With $\tilde x = x$ and $\tilde \pi_{i} = (R^{-1})_{i}^{j} \pi_{j}$ we have a (super)canonical transformation.

The quantization of the fermion oscillator is somewhat subtle, due to the fact that it is a constrained Hamiltonian system. In particular, it has the (second class) constraints $\pi_{i} + (i/2)\delta_{ij}\theta^{j} = 0$. These can be used to eliminate the momenta (on shell). Because of the constraints naive canonical quantization cannot be applied directly using the Poisson bracket, but may be implemented via the Dirac bracket \cite{Henneaux:1992ig} or the Peierls bracket \cite{DeWitt:1992cy}, leading to the (equal time) operator anticommutator
\begin{equation}
\{ \hat \theta^{i}, \hat \theta^{j} \} = \delta^{ij}.
\end{equation}
From these we can define an annihilation operator $a = \frac{1}{\sqrt 2} (\hat\theta^{1}- i\hat\theta^{2})$ and its conjugate creation operator $a^{\dagger}$, in terms of which we can write the Hamiltonian: $H = \omega(a^{\dagger}a- \frac 12)$. The spectrum of $H$ is $\{-\omega/2,+\omega/2\}$ and the non-degenerate Hilbert space is two-dimensional. If we have $N$ fermion oscillators then the Hilbert space is of dimension $2^{N}$.

We can discuss corrections to the strictly adiabatic case that are made manifest by the above supercanonical transformation, in parallel with our discussion of the bosonic case. In the transformation $x$ enters through the dependence of $R(x)$.  Remember that $R$ is a matrix that block-diagonalizes $M$. If we have only one fermion, $R$ is trivial and the state with the fermion occupied and the fermion unoccupied do not mix. This can also be seen from conservation of fermion parity symmetry.

In general we get corrections in derivatives of $R$. 
For two fermions there are four states in the quantum theory, two with even parity and two with odd parity. The two with even parity form a two-level system with a similar Hamiltonian to the one we computed in \eqref{eq:2level}, except that the off-diagonal terms are not just $(\Omega'/\Omega) v$, but involve the components of the $\partial R/\partial x$ matrix. These velocity corrections are qualitatively similar to those for bosons truncated to a two level system: after diagonalizing we will get a square root formula with momentum dependence for the zero point energy. Notably the correction to the zero point energy carries the same (negative) sign as in the bosonic case.

If there is no mixing between different fermions because $M$ is strictly block diagonal for all $x$, then we find that there is no fermionic contribution to the momentum-dependent forces, nor any fermion ``particle production'' unless $M$ changes sign (this is what happens in the simplified model studied by \cite{Douglas:1996yp}). Similar effects were found in \cite{Parker:1971pt}.

\section{Abrupt Changes}
\label{fast}

We can also consider non-adiabatic situations. In general the non-adiabatic case is not solvable and one has to resort to simulations. Here we consider a simple example that is tractable so that we can extract lessons for more general cases.

\begin{figure}
\centering
	\includegraphics[width=3in]{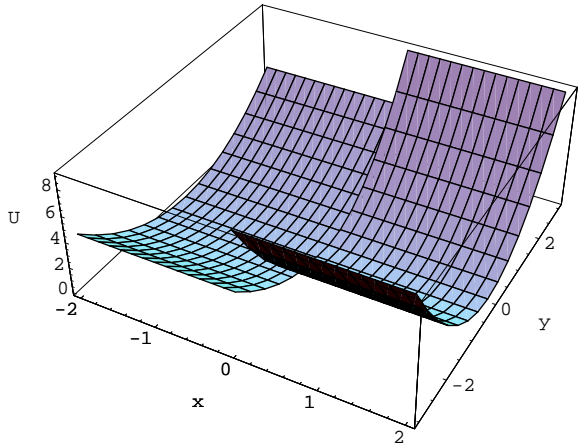}
	\caption{The abruptly changing potential}
	\label{pot1}
\end{figure}

  A particularly simple and extreme example involves an abruptly changing potential where 
\beq
	\Omega^2(x)  = \left\{
     \begin{array}{lr}
       1 & : x \leq 0 \\
       \omega^2 & : x >0,
     \end{array}
   \right.
\eeq
with $\omega^2 >0$ and $\omega \neq 1$. In this setup the wave functions on the left $x<0$ and right $x>0$ are solvable as they reduce to two separable degrees of freedom. The system reduces to solving 
a boundary condition. Notice that in this case we can always choose the mass of the degree of freedom $x$ to be equal to one by rescaling. We expect both reflection and transmission from such a potential, which is depicted in figure \ref{pot1}.

The physical situation we are modeling is a particle coming in from $x = -\infty$ and being scattered by the change in the potential. Alternatively, this represents an object with two degrees of freedom, one of which represents some center of mass motion (the $x$ variable) and another that represents some internal excitation (the $y$ variable)
whose structure changes at $x=0$.

To solve this problem, we solve the Schr\"{o}dinger equation away from the origin, imposing boundary conditions on the solution appropriate to the physical situation described above. One is free to choose an incoming state. We impose that the incoming particle has momentum $k$ and that it is in the ground state with respect to the $y$ variable, hence it has energy $E = k^2/2 + 1/2$.  The classical limit is where the energy stored in the $x$ degree of freedom is very large, expressed by a large $k$. 

Left of the origin (i.e. $x < 0$), we have the normalized solution
\begin{align}
	\psi(x,y) = A \left(\frac{1}{\pi}\right)^{1/4} e^{-y^2/2} e^{ikx} 
	+ \sum_{l=0}^\infty B_l \left(\frac{1}{\pi}\right)^{1/4} \frac{1}{\sqrt{l! 2^l}} H_l (y) e^{-y^2/2} e^{-i k_l x}.
\end{align}
The sum over modes is the set of reflected waves with excitations of the $y$ degree of freedom.

Conservation of energy requires that $E = k^2/2 + 1/2 = k_l^2/2 + l + 1/2$  for each $l$, or 
\beq
k_l = \sqrt{k^2 - 2l}.
\eeq
We always take the positive branch of the square root, ensuring that for $l < k^2/2$ we have a propagating mode to the left and for $l > k^2/2$ we have a decaying mode to the left. If this were a field theory problem, we would need to match the energy with zero point subtraction, which would be the same since both of them are to the left.

	In the positive $x$ region we can similarly write the wave function as
\begin{align}
	\psi(x,y) &= \sum_{l_1=0}^\infty C_{l_1} \left(\frac{\omega}{\pi}\right)^{1/4} \frac{1}{\sqrt{l_1! 2^{l_1}}} H_{l_1} (\sqrt{\omega}y) e^{-y^2\omega/2} e^{i k'_{l_1} x},
\end{align}
with conservation of energy here requiring $E = k^2/2 + 1/2 = {k'_{l_1}}^2/2 + \omega(l_1 + 1/2)$, or
\beq
	k'_{l_1} = \sqrt{k^2 + 1 - \omega(2l_1 + 1)},
\eeq
again taking the positive branch so we have propagation or decay to the right. We choose no incoming waves from the right. 

In a field theory setup the zero point energy on the right and left would be different and in general it would be infinite. To match we would need to work out a renormalized model properly taking into account the finite parts of the zero point energy. Here we can model it with an offset
in the conservation of energy if we want to. Only the conservation of energy equation for matching between left and right would require such an offset. In this case it would represent an additional contribution to the potential in the $x$ direction given by $\alpha \theta(x)$. In what follows we will not consider this possibility further: the results are similar to those with $\alpha=0$.

Our goal now is to calculate $B_l$ and $C_{l_1}$ exactly. These coefficients represent an S-matrix type problem between many channels meeting at a boundary. We must impose continuity of $\psi$ and $\partial \psi/\partial x$ at $x = 0$. To accomplish this, we need to express $H_n(\sqrt{\omega}y)$ in terms of $H_n(y)$ via a transformation matrix, $U_{mn}$. This can be derived directly from the generating function for the Hermite polynomials. It can also be computed from the creation and annihilation algebras as well as the corresponding vacuum states in the two regions, related by a Bogoliubov transformation. We describe this latter calculation in Appendix 1, where we also describe some further properties of $U_{mn}$. We find that 
\begin{align}
\label{U1}
	U_{mn} &:= \left(\frac{1}{\pi}\right)^{1/4} \left(\frac{\omega}{\pi}\right)^{1/4} \frac{1}{\sqrt{m!2^m}} \frac{1}{\sqrt{n!2^n}} \int_{-\infty}^{\infty} H_n(x) H_m(\sqrt{\omega}x) e^{-x^2(1+\omega)/2}\ dx \\
	\nonumber &= \sqrt{\frac{m!n!}{2^m 2^n}} \sqrt{\frac{2\omega^{1/2}}{1+\omega}}\frac{1}{((m+n)/2)!} \\
	\label{U2} &\times \sum_{k=0}^{(m+n)/2} {(m+n)/2 \choose k, k + (m-n)/2, n - 2k} \alpha^k (-\alpha)^{k+ (m-n)/2} \beta^{n-2k},
\end{align}
where the first factor in the sum is a trinomial coefficient ${a \choose b, c, d} = \frac{a!}{b!c!d!}$, and
\bea
	\alpha &:= \frac{1-\omega}{1+\omega} \\
	\beta &:= \frac{4\omega^{1/2}}{1+\omega}.
\end{align}
In the formula we must also add that the integral is zero unless $m$ and $n$ have the same parity. By orthonormality of the Hermite polynomials (properly normalized and weighted), we may write 
\beq
	\left(\frac{\omega}{\pi}\right)^{1/4} \frac{1}{\sqrt{m!2^{m}}} H_{m}(\sqrt{\omega} y) e^{-y^2\omega/2} = \sum_{n=0}^\infty U_{mn}  \left(\frac{1}{\pi}\right)^{1/4} \frac{1}{\sqrt{n!2^n}} H_n(y) e^{-y^2/2},
\eeq 
or in terms of the harmonic oscillator stationary wave functions
\beq
	u_n (y,\omega) := \left(\frac{\omega}{\pi}\right)^{1/4} \frac{1}{\sqrt{n!2^n}} H_n(\sqrt{\omega}x) e^{-y^2\omega/2}
\eeq
we have
\beq
	u_m (y,\omega) = \sum_{n=0}^\infty U_{mn} u_n (y,1).
\eeq
Note the matrix $U_{mn}$ is a change of basis matrix between two real orthonormal bases, and as such is orthogonal.

Now we can impose continuity of $\psi$ at $x = 0$. Approaching from the negative $x$ region, we have 
\beq
\label{psil}
	\psi(0, y) = A u_0(y,1) + \sum_{l=0}^\infty B_l u_l(y,1)
\eeq
and from the positive $x$ region
\bea
	\nonumber \psi(0,y) &= \sum_{l_1=0}^\infty C_{l_1} u_{l_1}(y,\omega) \\
	\nonumber &= \sum_{l_1=0}^\infty  C_{l_1} \sum_{l_2=0}^\infty U_{l_1 l_2} u_{l_2}(y,1) \\
	\label{psir}
	&= \sum_{l_1, l_2 = 0}^\infty C_{l_1} U_{l_1 l_2} u_{l_2}(y,1).
\end{align}
Now we can match the coefficients of the $u_l$ in \eqref{psil} and \eqref{psir}, which we can write as a matrix equation:
\beq
\label{psicont}
	\left( \begin{array}{ccc} A & 0 & \ldots \end{array} \right) + \left( \begin{array}{ccc} B_0 & B_1 & \ldots \end{array} \right) = \left( \begin{array}{ccc} C_0 & C_1 & \ldots \end{array} \right)
\left( \begin{array}{ccc}
U_{00} & U_{01} & \dots \\
U_{10} & U_{11} & \ \\
\vdots & \  & \ddots \end{array} \right),
\eeq
or more succinctly $A + B = CU$, with obvious matrix notation.

Similarly, we can impose the continuity condition on the derivative $\partial \psi/\partial x$. Coming from the negative $x$ region we have
\beq
\frac{\partial \psi}{\partial x} (0,y) = ikA u_0(y,1) - \sum_{l=0}^\infty ik_l B_l u_l(y,1)
\eeq
and from the positive $x$ region
\bea
	\nonumber \frac{\partial \psi}{\partial x} (0,y) &= \sum_{l_1=0}^\infty i k'_{l_1} C_{l_1} u_{l_1}(y,\omega) \\
	&= \sum_{l_1, l_2 = 0}^\infty i k'_{l_1} C_{l_1} U_{l_1 l_2} u_{l_2}(y,1).
\end{align}
So then the continuity of  $\partial \psi/\partial x$ at $x=0$ can be written as 
\beq
\label{dpsicont}
	\left( \begin{array}{ccc} kA & 0 & \ldots \end{array} \right) - \left( \begin{array}{ccc} k_0 B_0 & k_1 B_1 & \ldots \end{array} \right) = \left( \begin{array}{ccc} k'_0 C_0 & k'_1 C_1 & \ldots \end{array} \right)
\left( \begin{array}{ccc}
U_{00} & U_{01} & \dots \\
U_{10} & U_{11} & \ \\
\vdots & \  & \ddots \end{array} \right).
\eeq
If we define the matrices 
\beq
	{k} := \left( \begin{array}{ccc}
k_0 & 0 & \dots \\
0 & k_1 & \dots \\
\vdots & \vdots & \ddots \end{array} \right)
\eeq
and
\beq
	k' := \left( \begin{array}{ccc}
k'_0 & 0 & \dots \\
0 & k'_1 & \dots \\
\vdots & \vdots & \ddots \end{array} \right),
\eeq
then we can write \eqref{dpsicont} succinctly as $Ak - B {k} = Ck'U$.

To solve the scattering problem (getting $B$ and $C$ in terms of $A$), we must solve \eqref{psicont} and \eqref{dpsicont} together. Solving for $C$, we get
\beq
\label{Csoln}
	C = 2 A(k )(Uk + k'U)^{-1}.
\eeq
Solving for $B$ we get
\beq
\label{Bsoln}
	B = A (k - U^{-1}k' U)({k} +  U^{-1} k' U)^{-1}.
\eeq
This gives us the S-matrix for this potential.

	We can analyze these solutions by computing the corresponding transmission and reflection coefficients. These are usually defined for one-dimensional scattering problems in terms of the probability flux in the $x$ direction, which we can also do here. Recall that the probability current vector $j$ is defined by $j_i := \frac{-i}{2} \left(\psi^*\partial_i \psi - \psi \partial_i \psi^* \right)$ and its conservation is ensured by the Schr\"odinger equation. We can then use Stoke's theorem on a region $C$ bounded by some lines at $x = \pm c$ to obtain
\begin{align}
	0 = \int_C \nabla\cdot j(x,y)\ dx\ dy &= \int_{-\infty}^{\infty} j(-c,y)\cdot (-\hat{x})\ dy + \int_{-\infty}^{\infty} j(c,y)\cdot (\hat{x})\ dy  \\
	&=-k \abs{A}^2 + \sum_l \Real(k_l) \abs{B_l}^2 + \sum_{l_1} \Real(k'_{l_1}) \abs{C_{l_1}}^2.
\end{align}
Identifying the total transmission coefficient 
\beq
	T = \frac{\sum_{l_1} \Real(k'_{l_1}) \abs{C_{l_1}}^2}{k \abs{A}^2}
\eeq
and the total reflection coefficient
\beq
	R = \frac{\sum_l \Real(k_l) \abs{B_l}^2}{k \abs{A}^2}
\eeq
we have $T + R = 1$. We can also define the partial transmission and reflection coefficients for a given level $l$ as $T_l =  \Real(k'_{l}) \abs{C_{l}}^2/k \abs{A}^2$ and $R_l =  \Real(k_l) \abs{B_{l}}^2/k \abs{A}^2$.

\begin{figure}
\centering
	\includegraphics[width=4in]{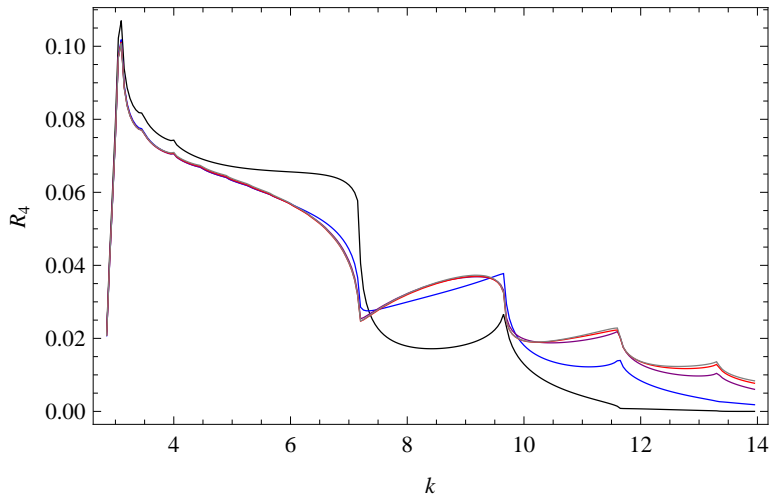}
	\caption{Plot of the partial reflection $R_4$ for $\omega = 10.5$ and matrices of size $10\times 10$, $20\times 20$, $40\times 40$, $60\times 60$, and $80 \times 80$,  shown in black, blue, purple, red, and gray respectively.}
	\label{Ndepend1}
\end{figure}

\begin{figure}
\centering
	\includegraphics[width=4in]{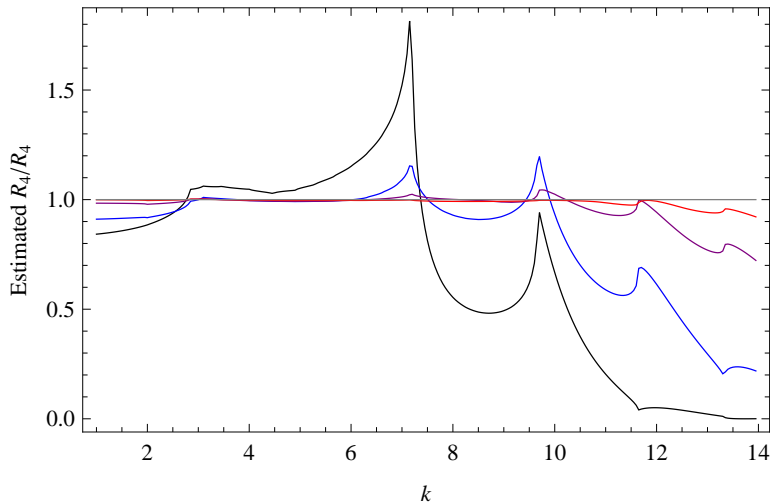}
	\caption{Plot of the relative discrepancy ($R_{4}$ with $N\times N$ truncation)/($R_{4}$ with $80\times 80$ truncation) for $\omega = 10.5$ and $N = 10, 20, 40, 60$ and $80$, shown in black, blue, purple, red, and gray respectively.}
	\label{Ndepend2}
\end{figure}

	We can calculate these coefficients for a given incoming amplitude using \eqref{Csoln} and \eqref{Bsoln}, but we run into the problem of inverting certain infinite matrices. We do this numerically by truncating the matrices to some finite size. We also ensure that the truncated version of the matrix $U$ in \eqref{U1} is unitary by orthogonalizing (and normalizing) its columns via the Gram-Schmidt algorithm. This allows us to avoid numerical issues encountered because simply truncating $U$ to finite size results in a matrix with very small eigenvalues. As the orthogonalized truncation is taken to large matrices, our results should converge to their true values. We have checked this numerically as shown in Figure \ref{Ndepend1}. There we show the curve for the second partial reflection coefficient for $\omega =10.5$ for matrices of increasing size. We see that the curve converges rather rapidly. To separate the curves better, we also show the relative discrepancy from the value calculated with $80\times 80$ matrices in Figure \ref{Ndepend2}. The lesson from numerous such numerical investigations is that, expectedly, one needs matrix sizes much larger than the highest channel one wants to accurately consider. 
	
\begin{figure}
\centering
	\includegraphics[width=5in]{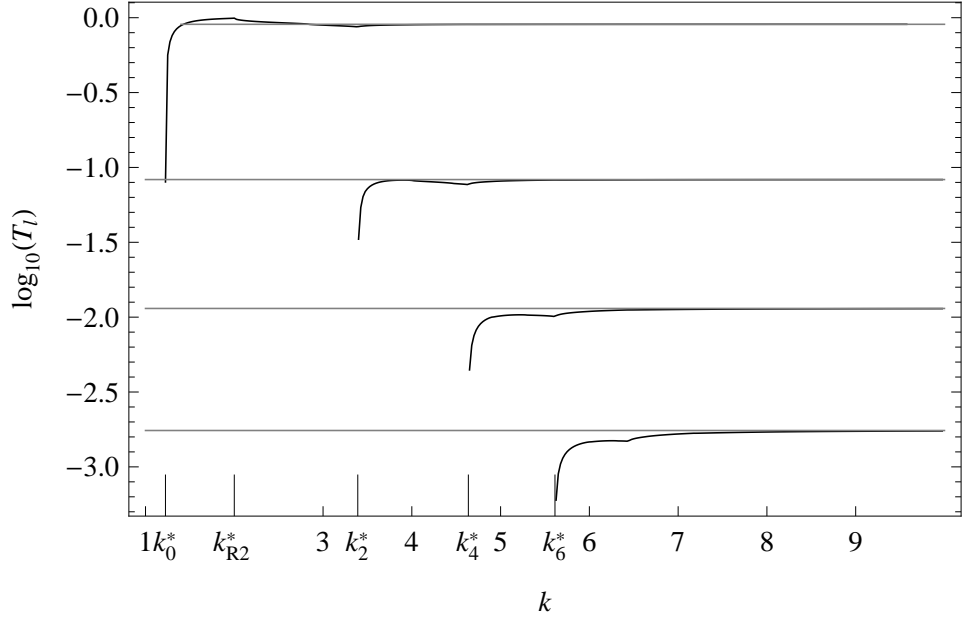}
	\caption{Logarithmic plot of the first four partial transmission coefficients for the case $\omega = 2.5$.}
	\label{trans1}
\end{figure}

	In Figure \ref{trans1} we display a logarithmic plot (base 10) of the first four partial transmission coefficients for the case $A = 1, \omega = 2.5$, for $k$ values ranging from $1.0$ to $10.0$ and calculated using  $50\times 50$ matrices. On the horizontal axis we show the values of momenta $k^*_l = \sqrt{\omega(2l +1) -1}$ at which it becomes kinematically possible to transmit into channel $l$. As expected, this is where we see the corresponding partial transmission coefficient become non-zero. 
	
	Also shown are the asymptotic values expected for these coefficients in the limit of large $k$. These come from the high $k$ limit of \eqref{Csoln}, which is $C = AU^{-1}$. In this limit there is no reflection and it corresponds to the classical limit for $x$. This leads to $T_l \rightarrow U_{l0}^2$, as shown. This is the result one would expect from just performing a Bogoliubov transformation between the mode functions for a time dependent 
abrupt change in a potential. We expect these asymptotic values to decrease exponentially with level number (See Appendix 1). The proof that this is the correct result is that asymptotically we have that $k_l \simeq k'_{l} \simeq k_0$ for all modes, up to corrections of order $1/k$. Thus $Uk' U^{-1} \simeq k$ and $B_n\to 0$.

\begin{figure}
\centering
	\includegraphics[width=5in]{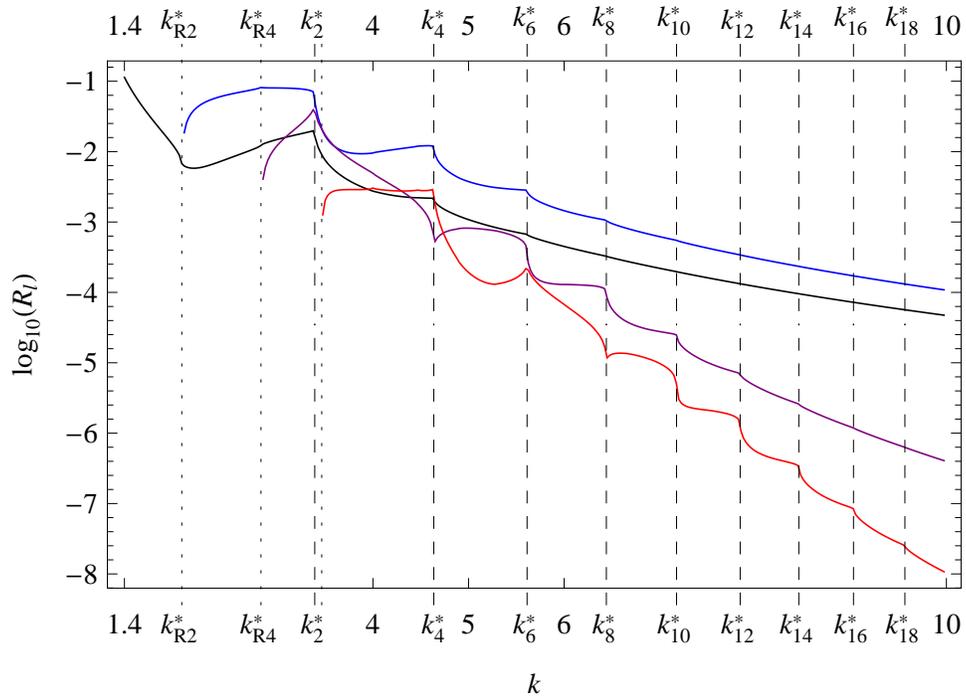}
	\caption{Logarithmic Plot of the first four partial reflection coefficients for the case $\omega = 2.5$. $R_0, R_2, R_4, R_6$ are shown in black, blue, purple, and red, respectively.}
	\label{refl1}
\end{figure}

Notice that at finite $k$ values the conservation of energy plays an important role in forbidding transitions to high excitation. For these transitions the naive classical result of the Bogoliubov transformation gives the wrong answer and the probabilities converge at different values of $k$ for each mode. Our plots show how that works. The other thing to notice is that reflection coefficients are generally small. However, these would not be present at all in classical physics.
	
	The logarithmic plot of the partial reflection coefficients contains more structure, as shown in Figure \ref{refl1}. As expected the coefficients all approach zero in the limit of high $k$, corresponding to negative infinity on the plot. We have restricted the range of $k$ to $(1.4,10.0)$ to increase the contrast. The vertical dashed lines correspond to the onset of transmission at momenta $k^*_l$ as above. The vertical dotted lines indicate when a new reflection channel opens up, $k^*_{Rl} = \sqrt{2l}$.
	
	We would like to remind the reader that having reflection in this setup is a purely quantum effect and can not be obtained from a semiclassical expansion around a classical trajectory for $x$. Because we have solved the problem exactly, as given by equations \eqref{Csoln} and \eqref{Bsoln}, our system has taken into account back-reaction to the quantum $y$ motion exactly. We also can see that the system converges to the usual treatment of a time dependent harmonic oscillator in the classical limit for $x$ and we can quantify the convergence precisely. The conservation of energy also tells us that each mode $k'_l$ will propagate at different speeds, so that the different components of the wave function will separate. This effect can be measured and further interactions with other degrees of freedom would probably decohere these different branches of the wave function.

In a similar problem in field theory, the mass $m_x$ is related to the volume of space.  The large mass limit is equivalent to the large volume limit. One can also state that the large mass limit is equivalent to a weak coupling limit by a rescaling of fields associated with $x$, but this does not mean much in this abrupt scenario.  Notice that one gets finite particle production per unit volume and the general matching involves all the modes of the heavy fields. Taking the large volume limit is tricky because the $x$ degree of freedom becomes part of a field. The right way to solve the problem is to coarse grain it into cells
of fixed volume, so that gradients between adjacent cells are subleading in energy and so that the wavelengths of the particles produced are typically much shorter than the size of the cells. In such a way one can in principle perform the above calculation cell by cell and have the possibility of reflection  or not on different boxes generating fluctuations in $x$.  Notice that different positions will evolve afterwards at different rates (if one imposes energy conservation cell by cell), so the problem becomes rather involved even in the presence of a simplified abrupt potential as we have considered above.
Such a calculation is beyond the scope of the present paper.

%

\section{Conclusion}

	 From the simple system described by \eqref{H}, for one modulus $x$ and one harmonic oscillator (fast) degree of freedom whose frequency, $\Omega(x)$, depends on $x$,  we have gotten insight into the near-adiabatic and non-adiabatic dynamics of quantum systems.  Following the method of Born and Oppenheimer we have been able to clearly see the transition from adiabatic to non-adiabatic behavior, including back-reaction, and to understand when the oscillator degrees of freedom cannot be integrated out. We implemented an improved adiabatic approximation that gave us a velocity dependent potential for the moduli motion. This accounted for some back-reaction effects. We saw how this back-reaction depended on the velocity of the modulus $x$ and the coupling between $x$ and the oscillator, as well as how it can be interpreted as modifying the metric on the moduli space. We also argued that for finite time evolutions one obtains an occupation number for the oscillator degrees of freedom that is a power series in the non-adiabaticity parameter. These effects are responsible for the velocity dependent corrections to the Hamiltonian when the oscillator degree of freedom is integrated out. However, for infinite time behavior the net particle production can be non-perturbatively small.

	 Despite the simplicity of the model, which we have analyzed in a non-relativistic classical and quantum mechanical setting, it is general enough to be relevant to both relativistic quantum field theory and string theory as we have briefly discussed and as evidenced by its frequent appearance (for various choices of $\Omega$) in the literature. We plan to apply these results to these setups in the future. In particular, we found that the velocity corrections have a definite sign, the same for fermions and bosons. However, we did not carry out a complete analysis for gauged systems, which may be behave differently in important ways (e.g. because of the presence of ghosts). This is especially important to address because of the non-renormalization theorems that constrain the dynamics, via cancellations, when supersymmetry is present \cite{PSS}. It is important to understand exactly how these cancellations happen in supersymmetric gauged systems, as in those systems associated with D-branes.

	 We have also solved a fully non-adiabatic problem exactly and investigated the solution numerically. Our model had an abrupt change in $\Omega$ that gave rise to quantum mechanical transmission and reflection, as in one-dimensional scattering problems, which are displayed in the final four figures. The kinematics gave rise to a separation of different modes, leading to possible decoherence of the wavefunction amplitudes as time goes on and the system interacts with other degrees of freedom. It is easy from our solution to treat the interesting case of an $\Omega$ with multiple abrupt changes and we plan to investigate this in future work. 
	 
	 Many of the phenomena we have encountered are inherently quantum-mechanical and may be important to consider in applications, such as D-brane interactions or inflationary cosmology, where previous analyses have been semiclassical in nature. It may be crucial in making a problem tractable that these effects are suppressed, so that one may get by with semiclassical physics. We think it is thus important to examine very carefully when and to what extent this is the case. We found the above model, which admitted exact quantum solutions in certain simple situations, an illuminating test case for such an investigation.

\section*{Acknowledgements}

We would like to thank E. Silverstein for various discussions.
C. A. thanks I. Heemskerk, B. Horn and M. Roberts for several helpful discussions and TASI 2010 for its hospitality while working on this paper. D. B. would like to thank M. Douglas, J. Hartle and E. Verlinde for various discussions related to this work. D. B. would also like to thank the Simons Center for Geometry and Physics for their hospitality during the final stages of this work. D. B.  is supported in part by the DOE under grant DE-FG02-91ER40618.

\section{Appendix 1}
\label{Bog}

	The Hamiltonian for the $y$ variable in the two regions separated by the $y$-axis differ only in their value of $\Omega$. Let us call the region where $x<0$ region 1 and where $x>0$ region 2 and consider a Hamiltonian with two different values for $\Omega$, $\omega_1$ and $\omega_2$ respectively. Since we are interested just in ensuring continuity of the wavefunction at $x=0$ we can ignore the $x$ dependence of the states here. The problem then is to find the transformation that will take us from the basis of energy eigenstates, which we can write as $\{|n\ket_1\}$, in one region to the basis $\{|n\ket_2\}$ in the other. That is, we want to calculate the matrix in \eqref{U1}:
\beq
	U_{mn} = \bra n |_1 m \ket_2 = \frac{1}{\sqrt{n! m!}} \bra 0 |_1 (a_1)^n (a^\dagger_2)^m |0\ket_2.
\eeq
This can be accomplished by relating the creation and annihilation operators as well as the vacuum states on the two sides. For $i = 1, 2$ we have
\begin{align}
	a_i &= \sqrt{\frac{\omega_i}{2}}\left(y + \frac{i p_y}{\omega_i} \right) \\
	a^\dagger_i  &= \sqrt{\frac{\omega_i}{2}}\left(y - \frac{i p_y}{\omega_i} \right).
\end{align}
Then we can calculate
\beq
	\begin{pmatrix} a_1 \\ a^\dagger_1 \end{pmatrix} = \frac{1}{2\sqrt{\omega_1 \omega_2}} \begin{pmatrix} \omega_1 + \omega_2 & \omega_1 - \omega_2 \\ \omega_1 - \omega_2 & \omega_1 + \omega_2 \end{pmatrix} \begin{pmatrix} a_2 \\ a^\dagger_2 \end{pmatrix},
\eeq
a Bogoliubov transformation \cite{BD,PT} or quasiparticle transformation \cite{Merz}. To relate the vacuum states $|0\ket_i$ we note that they are squeezed states with respect to the opposite operator algebra. To find an explicit formula for $|0\ket_1$ we recall the representation of states by holomorphic functions \cite{Merz} and write $|0\ket_1 = F(a^\dagger_2) |0\ket_2$ for $F$ holomorphic. Then since formally $a_2 = \partial/\partial a^\dagger_2$ when acting on the vacuum state, $a_1 |0\ket_1 = 0$ yields the equation for $F$
\beq
	\left( \left(\omega_1 + \omega_2\right) \frac{\partial}{\partial z} + \left(\omega_1 - \omega_2\right)z \right) F(z) = 0,
\eeq
which has solution $F(z) = C \exp \left(-\frac{1}{2}\frac{\omega_1 - \omega_2}{\omega_1 + \omega_2} z^2 \right)$. Normalizing, we find
\beq
	|0\ket_1 = \left( \frac{2\sqrt{\omega_1\omega_2}}{\omega_1+\omega_2} \right)^{1/2} \exp\left( -\frac{1}{2} \frac{\omega_1-\omega_2}{\omega_1+ \omega_2} \left(a^\dagger_2\right)^2 \right) |0\ket_2.
\eeq
Thus
\begin{align}
\label{U3}
	\nonumber U_{mn} = &\frac{1}{\sqrt{n!m!}} \left( \frac{2\sqrt{\omega_1\omega_2}}{\omega_1+\omega_2} \right)^{1/2} \left(\frac{1}{\sqrt{\omega_1\omega_2}} \right)^n \\ 
	&\times \left\bra 0 \phantom{ \mathrm{\bigg \vert}} \right\vert_2 \ \exp\left( -\frac{1}{2} \frac{\omega_1-\omega_2}{\omega_1+ \omega_2} \left(a_2\right)^2 \right) \left[(\omega_1 + \omega_2)a_2 + (\omega_1 - \omega_2)a^\dagger_2 \right]^n \left(a^\dagger_2\right)^m \left\vert\phantom{ \mathrm{\bigg \vert}}0\right\ket_2.
\end{align}
This expresssion can in principle be evaluated using the commutation relations for $a_2$ and $a^\dagger_2$, though we found it easier to work directly with the Hermite polynomial generating function to derive \eqref{U2}. But \eqref{U3} is useful in determining the large $m$ or large $n$ asymptotics of the off-diagonal elements and hence of the partial transmission coefficients considered in section \ref{fast}. 

	First we note a certain symmetry property of the matrix $U_{mn}$, which is easiest to see from the integral expression \eqref{U1}. Since the functions $H_n(x) \exp(-x^2/2)$ are eigenfunctions of the (unitary) Fourier transform, with eigenvalue $(-i)^n$, Fourier transforming the functions in \eqref{U1} yields 
\begin{align}
	\nonumber U_{mn} &= (-i)^{m+n} \frac{1}{2\pi} \iiint u_n(p) u_m(p') e^{ix(p+p'\sqrt{\omega})} \ dx\ dp\ dp' \\
	\nonumber &=  (-i)^{m+n} \iint u_n(p) u_m(p') \delta(p+p'\sqrt{\omega})\  dp\ dp' \\
	&=  (-i)^{m+n} U_{nm}.
\end{align}
This says $U_{mn}$ is symmetric up to a phase.
	
	Since here we are just interested in the absolute value of the off-diagonal elements, by the above symmetry property we lose no generality if we fix $n$ and consider large $m$. The leading contribution comes from the term in the expansion of the exponential in \eqref{U3} with the appropriate power to annihilate the $a^\dagger_2$ operators. We take $m \gg n$ and focus on the $m$-dependent factors, using the Stirling approximation on all factorials. We find that the $m$ dependence is asymptotically
\beq
	U_{mn} \sim  \frac{1}{m^{1/4}} \left(- \frac{\omega_1-\omega_2}{\omega_1+ \omega_2} \right)^{m/2},
\eeq
which decreases exponentially in magnitude with $m$.

\end{document}